\documentclass{llncs}

\usepackage{graphicx}
\usepackage{listings}
\usepackage{todonotes}
\usepackage{xspace}
\usepackage{color}
\usepackage{array}
\usepackage{underscore}
\usepackage{amsfonts}
\usepackage{tipa}

\def\systemname#1{\textsf{#1}\xspace}

\newcommand{\Tactician}{\systemname{Tactician}}
\newcommand{\HOLLight}{\systemname{HOL Light}}
\newcommand{\HOL}{\systemname{HOL}}
\newcommand{\Isabelle}{\systemname{Isabelle}}
\newcommand{\Mizar}{\systemname{Mizar}}
\newcommand{\Vampire}{\systemname{Vampire}}

\newcommand{\Z}{\systemname{Z3}}
\newcommand{\E}{\systemname{E}}
\newcommand{\Coq}{\systemname{Coq}}
\newcommand{\Flyspeck}{\systemname{Flyspeck}}

\newcommand{\OCaml}{\systemname{OCaml}}
\newcommand{\Proviola}{\systemname{Proviola}}
\newcommand{\Agora}{\systemname{Agora}}

\newcommand{\MathJax}{\systemname{MathJax}}
\newcommand{\HH}{\systemname{HOL(y)Hammer}}

\usepackage{hyperref}

\newcommand{\demourl}{\url{http://mws.cs.ru.nl/agora_cicm/flyspeck/doc/fly_demo/}}

\makeatletter
\renewcommand\section{\@startsection{section}{1}{\z@}%
                       {-12\p@ \@plus -4\p@ \@minus -4\p@}%
                       {8\p@ \@plus 4\p@ \@minus 4\p@}%
                       {\normalfont\large\bfseries\boldmath
                        \rightskip=\z@ \@plus 8em\pretolerance=10000 }}
\makeatother

\begin{document}
\lstset{language=[Objective]Caml}

\title{Formal Mathematics on Display:\\ A Wiki for Flyspeck\thanks{The final publication is available at http://link.springer.com.}}
\author{Carst Tankink \and Cezary Kaliszyk \and Josef Urban \and Herman Geuvers}

\author {Carst Tankink\inst{1}
\and Cezary Kaliszyk\inst{2}
\and Josef Urban\inst{1}
\and Herman Geuvers\inst{1,3}
}
\authorrunning {C. Tankink
\and C. Kaliszyk
\and J. Urban
\and H. Geuvers
}
\institute{
ICIS,\ Radboud Universiteit Nijmegen, 
Netherlands
\and
Institut f\"ur Informatik, Universit\"at Innsbruck, Austria
\and
Technical University Eindhoven, Netherlands
}

\maketitle

\begin{abstract}
 The \Agora system is a prototype ``Wiki for Formal Mathematics'', with an aim
 to support developing and documenting large formalizations of
 mathematics in a proof assistant. The functions implemented in \Agora
 include in-browser editing, strong AI/ATP proof advice, verification,
 and HTML rendering. The HTML rendering contains hyperlinks and
 provides on-demand explanation of the proof state for each proof
 step.  In the present paper we show the prototype \Flyspeck Wiki as an instance
 of \Agora for \HOLLight formalizations. The wiki can be used for
 formalizations of mathematics and for writing informal wiki pages
 about mathematics. Such informal pages may contain islands of formal
 text, which is used here for providing an initial cross-linking
 between Hales's informal \Flyspeck book, and the formal \Flyspeck
 development.

The \Agora platform intends to address distributed wiki-style
collaboration on large formalization projects, in particular both the
aspect of immediate editing, verification and rendering of formal
code, and the aspect of gradual and mutual refactoring and
correspondence of the initial informal text and its
formalization. Here, we highlight these features within the \Flyspeck
Wiki.
\end{abstract}

\section{Introduction}
The formal development of large parts of mathematics is gradually
becoming mainstream.
In various proof assistants, large repositories of formal proof have
been created, e.g.\ in \Mizar~\cite{mizar-in-a-nutshell}, \Coq~\cite{BC04},
\Isabelle~\cite{nipkow-et-al-2002} and \HOLLight~\cite{Harrison-2009}. This has led
to fully formalized proofs of some impressive results, for example the
odd order theorem in \Coq~\cite{DBLP:conf/popl/Gonthier13}, the proof
of the 4 color theorem in \Coq~\cite{DBLP:conf/ascm/Gonthier07} and
a significant portion of the proof of the Kepler
conjecture~\cite{DBLP:journals/dcg/HalesHMNOZ10} in \HOLLight.

Even though these results are impressive, it is still quite hard to
get a considerable speed-up in the formalization process. If we look
at Wikipedia, we observe that due to its distributed nature everyone
can and wants to contribute, thus generating a gigantic increase of
volume. 
If we look at the large formalization projects, we see that they are
very hierarchically structured, even if they make use of systems like
Coq, that very well support a cooperative distributed way of working,
supported by a version control system. An important reason is
that the {\em precise\/} definitions {\em do\/} matter in a computer
formalised mathematical theory: some definitions work better than
others and the structure of the library impacts the way you work with
it. 

There are other reasons why formalization is progressing at a much
slower rate than, e.g.\ Wikipedia. One important reason is that it is
very hard to get access to a library of formalised mathematics and to
reuse it: specific features and notational choices matter a lot and
the library consists of such an enormous amount of detailed formal
code that it is hard to understand the purpose and use of its
ingredients. A formal repository consists of computer code
(in the proof assistant's scripting language), and has the same challenges as a programming source code regarding understanding,
modularity and documentation. Also, if you want to make a
contribution to a library of formalized mathematics, it really has to
be all completely verified until the final proof step. And finally, giving
formal proofs in a proof assistant is very laborious, requiring a significant amount of training and experience to do effectively.

To remedy this situation we have been developing the \Agora\ platform: wiki
technology that supports the development of large coherent
repositories of formalised mathematics. We illustrate our work by
focusing on the case of a wiki for the \Flyspeck project, but the aims
of \Agora\ are wider. In short we want to provide proof assistant users
with the tools to

\begin{enumerate}
\item Document and display their developments for others to be read
  and studied,
\item Cooperate on formalizations,
\item Speed up the proving by giving them special proof support via
  AI/ATP tools.
\end{enumerate}
All this is integrated in one web-based framework, which aims at being
a ``Wiki for Formal Mathematics''. In the present paper we highlight
and advocate our framework by showing the prototype \Flyspeck Wiki.
We first elaborate on the three points mentioned above and
indicate how we support these in \Agora.

\paragraph{Documenting formal proofs}
An important challenge is the communication of large
formalizations to the various different communities interested in such formalizations: PA users that want to cooperate or want to build further on the
development, interested readers who want to understand the precise
choices made in the formalization and mathematicians who want to
convince themselves that it is really the proper theorem that has been
proven. All these communities have their own views on a formalization and the process of creating formalization, giving a diverse input that benefits the field. Nonetheless, communicating a formal proof is hard, just as hard as communicating a computer program.

\Agora\ provides a wiki based approach: Formal proofs are basically
program code in a high-level programming language, which needs to be
documented to be understandable and maintainable. A proof
development of mathematics is special, because there typically is
documentation in the form of a mathematical text (a book or an
article) that describes the mathematics informally.  This is what we call the \emph{informal mathematics} as opposed to the \emph{formal mathematics} which is
the mathematics as it lives inside a proof assistant. For software verification efforts, there is no pre-existing documentation, but \Agora can be used to provide documentation of the verification as well. 
These days, informal mathematics consists of \LaTeX\ files and formal mathematics
usually consist of a set of text files that are given as input to a
proof assistant to be checked for correctness. 

In \Agora, one can automatically generate HTML files from formal proof
developments, where we maintain all linking that is inherently available in the
formal development. Also, one can automatically generate files in wiki syntax
from a set of \LaTeX\ files. These wiki files can also be rendered as HTML, maintaining the linking inside the \LaTeX\ files, but
more importantly, also the linking with the formal proof development.
Starting from the other end, one can write a wiki document about
mathematics and include snippets of formal proof text via an
inclusion mechanism. This allows the dynamic insertion of pieces of
formal proof, by referencing the formal object in a repository.

\paragraph{Cooperation on formal proofs} With \Agora, we also want to
lower the threshold for participating in formalization projects by
providing an easy-to-use web interface to a proof
assistant~\cite{Tankink-2012}. This allows people to cooperate on a project, the files of which are stored on the server.

\paragraph{Proof Support} 
We provide additional tools for users of proof assistants, like
automated proof advice~\cite{hhjournalarxiv}. The proof states
resulting from editing \HOLLight code in \Agora are
continuously sent to an online AI/ATP service which is trained in a
number of ways on the whole \Flyspeck corpus. The service automatically
tries to discharge the proof states by using (currently 28) different
proof search methods in parallel, and if successful, it attempts to
create the corresponding code reconstructing such proofs in the user's
\HOLLight session.

\paragraph{}To summarize, the \Agora system now provides the following tooling for \HOLLight and \Flyspeck:
\begin{itemize}
  \item a rendering of the informal proof texts, written originaly in \LaTeX,
  \item a hyperlinked, marked up version of the \HOLLight and
    \Flyspeck source code, augmented with the information about the
    proof state after each proof step
  \item transclusion of snippets of the hyperlinked formal code into
    the informal text whenever useful
  \item cross-linking between the informal and formal text based on
    custom \Flyspeck annotations
  \item an editor to experiment with the sources of the proof by
    dropping down to \HOLLight and doing a formal proof,
  \item integrated access to a proof
    advisor for \HOLLight that helps (particularly novices)
    to finish their code while they are writing it, or provide options for improvement, by suggesting lemmas that will solve smaller steps in one go.
\end{itemize}

Most of these tools are prototypical and occasionally behave in unexpected ways. The wiki pages for \Flyspeck can be found at \url{http://mws.cs.ru.nl/agora_cicm/flyspeck}. These pages also list the current status of the tooling.

The rest of the paper is structured as
follows. Section~\ref{sec:presentation} shows the presentation side of
\Agora, as experienced by readers. The internal document model of
\Agora is described in Section~\ref{Documents},
Section~\ref{Interaction} explains the interaction with the formal
\HOLLight code, and Section~\ref{Informal} describes the inclusion of
the informal \Flyspeck texts in \Agora. Section~\ref{Conclusion}
concludes and discusses future work.

\subsection{Similar Systems} 
There are some systems that support mashing up informal documentation with computed information. In particular, \Agora shares some similarities with tools using the OMDoc~\cite{OMDoc} format, as well as the \systemname{IPython}~\cite{IPython} architecture (and \systemname{Sage}~\cite{Sage}, which uses \systemname{IPython} as an interface to computer algebra functionality).

OMDoc is mainly a mechanization format, but supports workflows that are similar to \Agora's, but differs in execution: OMDoc is a stricter format, requiring documents to be more structured and detailed. In particular, this requires its input languages, such as s\TeX, to be more structured. On the other hand, \Agora does not define much structure on the files its includes, rather extracting as much information as possible and fitting it in a generic tree structure. Because \Agora is less strict in its assumptions, it becomes easier to write informal text, freeing the authors of having to write semantic macros.

The \systemname{IPython} architecture has the concept of a \emph{notebook} which is similar to a page in \Agora: it is a web page that allows an author to specify 'islands' of Python that are executed on the server, with the results displayed in the notebook. \Agora builds on top of this idea, by having a collection of documents referring to each other, instead of only allowing the author of a document to define new islands.

\section{Presenting Formal and Informal Mathematics in \Agora}
\label{sec:presentation}

\Agora has two kinds of pages: fully \emph{formal} pages, generated from the sources of the development, and \emph{informal} pages, which include both markup and snippets of formal text.
To give readers, in particular readers not used to reading the syntax of a proof assistant, insight in a formal development, we believe that it is not enough to mark up the formal text prettily: 

\begin{itemize}
  \item there is little to no context for an  inexperienced reader to quickly understand what is being formalized and how: items might be named differently, and in a proof script, all used lemmas are presented with equal weight. This makes it difficult for a reader to single out what is used for what purpose;
  \item typically, the level of detail that is used to guide the proof assistant in its verification of a proof is too high for a reader to understand the essence of that proof: it is typically decorated with commands that are administrative in nature, proof steps such as applying a transitivity rule. A reader makes these steps implicitly when reading an informal proof, but they must be spelled out for a formal system. In the extreme, this means that a proof that is `trivial' in an informal text still requires a few lines of formal code;
  \item because most proof assistants are programmable, a proof in proof assistant syntax can have a different structure than its informal counterpart: proofs can be `packed' by applying proof rules conditionally, or applying a proof rule to multiple similar (but not identical) cases.
\end{itemize}

On the other hand, it is not enough to just give informal text presenting a formalization: without pointers to the location of a proof in the formal development, it is easy for a reader to get lost in the large amount of code. To allow easier navigation by a reader, the informal text should provide \emph{references} to the formal text at the least, and preferably include the portions of formal text that are related to important parts of the informal discussion.

By providing the informal documentation and formal code on a single web platform, we simplify the task of cross-linking informal description to formal text. The formal text is automatically cross-linked, and annotated with proper anchors that can also be referenced from an informal text. Moreover, our system uses this mechanism to provide a second type of cross-reference, which includes a formal entity in an informal text~\cite{Tankink+-2012}: these references are written like hyperlinks, using a slightly different syntax indicating that an inclusion will be generated. Normal hyperlinks can refer to concepts on the same page, the same repository, or on external pages.

These mechanisms allow an author of an informal text to provide an overview of a formal development that, at the highest level, can give the reader insight in the development and the choices made. Should the reader be interested in more details of the formalization, cross-linking allows further investigation: clicking on links opens the either informal concepts or shows the definition of a formal concept.

\begin{figure}
\includegraphics[width=\textwidth]{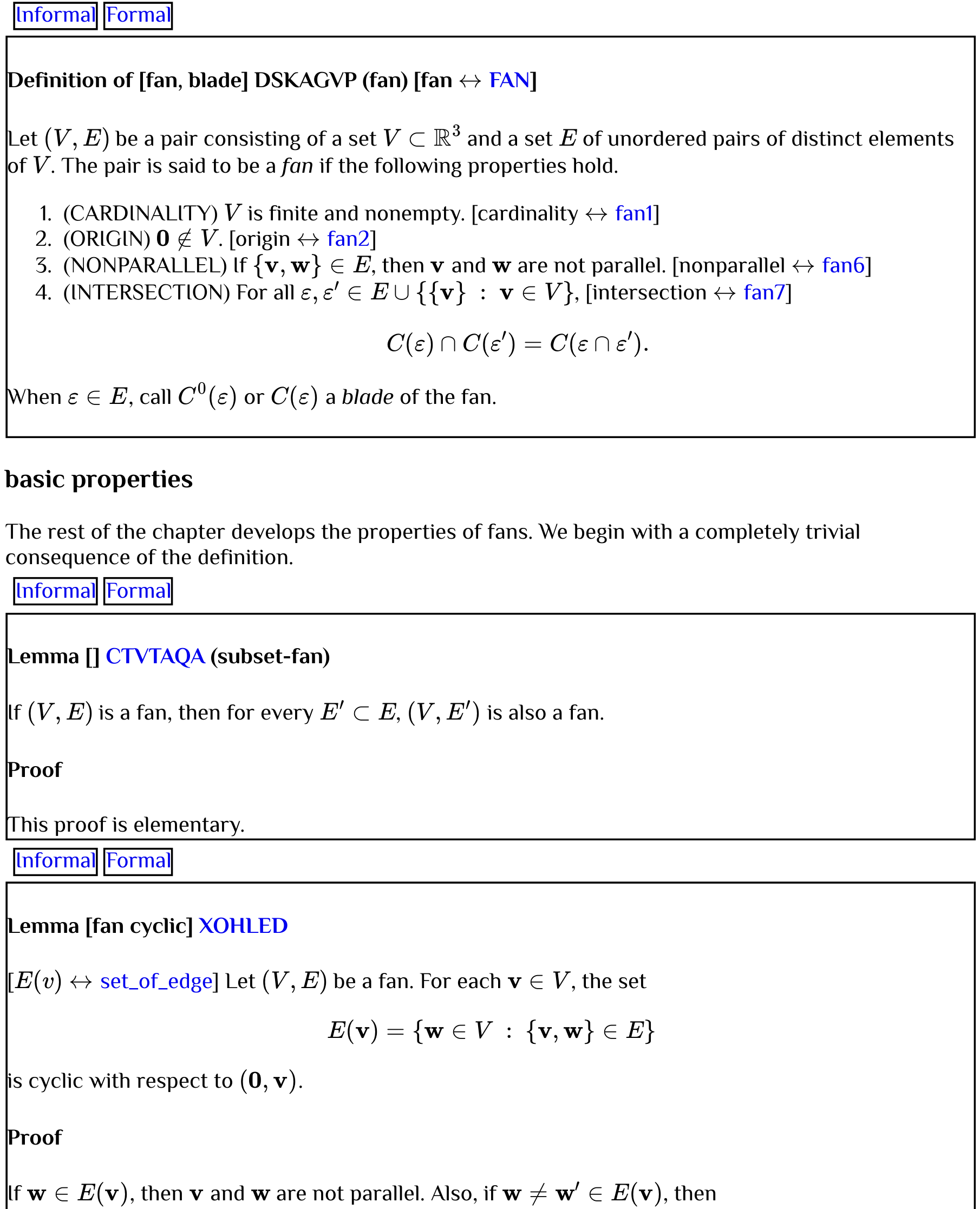}
\caption{Screenshot of the \Agora wiki page presenting a part of the ``Fan'' chapter of the informal
  description of the Kepler conjecture formalization. For each formalized section, the user can
  choose between the informal presentation (shown here) and its formal counterpart (shown
  on the next screenshot). The complete wikified chapter is available at: \demourl.}
\label{fig:screenshot2}
\end{figure}

\begin{figure}
\includegraphics[width=\textwidth]{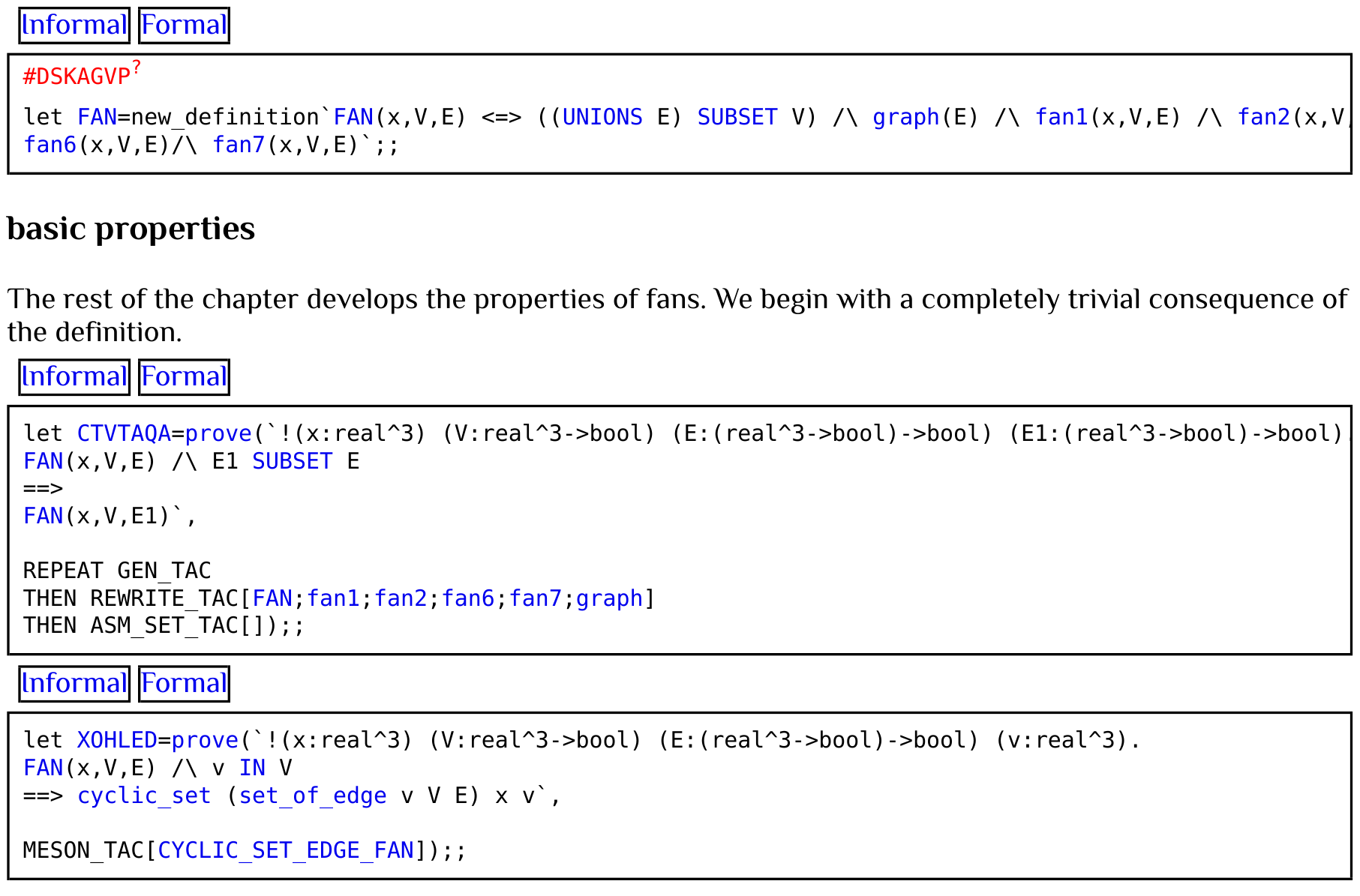}
\caption{\demourl (formal).}
\label{fig:screenshot}
\end{figure}

The formalization of the Kepler conjecture in the \Flyspeck project provides us with an opportunity to display these techniques: not only is it a significant non-trivial formalization, but its informal description in \LaTeX~\cite{Hales-2012} contains explicit connections between the informal mathematics and the related formal concepts in the development. We have transformed these sources into the wiki pages available on our \Agora system\footnote{\demourl}. Parts of one page are shown in Figures~\ref{fig:screenshot2} and \ref{fig:screenshot}. 

\subsection{Informal Descriptions}
The informal text on the page is displayed similarly to the source (\Flyspeck) document, from which it is actually generated (see Section~\ref{Creolifier}), keeping the formulae intact to be rendered by the \MathJax\footnote{\url{http://mathjax.org}} JavaScript library.
The difference to the \Flyspeck source document is that the source document contains \emph{references} to formal items (see also Section~\ref{Informal}), while the \Agora version \emph{includes} the actual text of these formal entities. To prevent the reader from being confused by the formal text, which can be quite long, the formal text is hidden behind a clearly-labeled link (for example the \texttt{FAN} and \texttt{XOHLED} links in Figure~\ref{fig:screenshot2} which link to the formal definition of \emph{fan} and the formal statement of lemma \emph{fan\_cyclic}). 

The informal page may additionally \emph{embed} editable pieces of formal code (instead of just including addressable formal entities from other files as done in the demo page). In that case (see Section~\ref{Interaction}) clicking the 'edit' on these blocks opens up an editor on the page itself, which gives direct feedback by calling \HOLLight in the background, and displaying the resulting proof assistant state, together with a \emph{proof advice} which uses automated reasoning tools to try to find a solution to the current goal.

\subsection{Formal Texts}
The formal text of the development, in the proof assistant syntax, is included in \Agora as a set of hyperlinked HTML pages that provide \emph{dynamic} access to the proof state, using the \Proviola~\cite{TankinkMcKinna-2011} technology we have previously developed: pointing at the commands in the formal text calls the proof assistant and provides the state on the page. The results of this computation are memoized for future requests: this makes it possible for future visitors to obtain these states quickly, while not taking up space unnecessarily.

The pages are hyperlinked (see Section~\ref{Hyperlinking}) to allow a reader to explore the presented formalization. The formalization could be large and, in projects like \Flyspeck, produced by a number of collaborators. The current alternatives to hyperlinking are unsatisfactory in such circumstances: it amounts to  either memorization by the reader of large parts of the libraries, or mandatory access to a search facility. In \HOLLight, this search facility is the system itself: typing in the name of a lemma prints out its statement.

\section{Document Structure: Frames and Scenes}
\label{Documents}
The pages in \Agora are generated from in-memory \emph{documents}: (Python) objects equipped with methods for rendering and storing the internal files. To cater for multiple proof assistants and document-preparation tools, such as a renderer for wiki syntax, we use the object-inheritance to instantiate documents for different systems, while providing a common interface. This interface consists of a tree-like structure of \emph{frames}, grouped into \emph{scenes}.

Documents in \Agora are structured according to our earlier work on a system called \Proviola~\cite{DBLP:conf/aisc/TankinkGMW10}, for replaying formal proof: this tool takes a ``proof script'' and uses a light-weight parser to transform it into a list of separate commands. This list can then be submitted to a proof assistant, storing the responses in the process. This memoization of the proof assistant's responses is stored together with the command, into a data structure we call a \emph{frame}. Frames can store more than just a response and a command, in particular, we assume that all frames in \Agora documents store a markup element that contains the HTML markup of the frame's command.

To display a document as a page, it would be enough to display the list of frames in order, rendering the markup of each frame, and this is how the purely formal pages in \Agora are rendered. However, we want our tools to be able to display not only flat lists of text, but also combine them in meaningful ways: for example by grouping a lemma with its proof, but also combining multiple lemmas into a self-contained section. For this, we introduced a \emph{scene}: a scene is a grouping of (references to) frames and other scenes, that can combine them in any order. The system will render such a tree structure recursively, displaying the markup of each frame referenced to. The benefit of grouping files into scenes is that it becomes easier to re-mix parts of a document into a new document, such as including formal text into an informal page.

\subsubsection{Inclusion}
To allow remixing scenes from documents into new content, it is necessary to provide an interface that allows including scenes into pages. In previous work~\cite{Tankink+-2012}, we introduced an interface in the form of syntax: \Agora allows users to write narratives in a markup language similar to \systemname{Wikipedia}'s, which is extended with the notion of a \emph{reference}. This reference is similar to \Isabelle's antiquotation: it is syntax for pointing to formally defined entities on the Web which carry some metadata, which can be automatically provided by a theorem prover. When rendered, the references are resolved into marked up `islands' of formal text. The rest of the syntax is a markup language allowing mathematical notation and hyperlinks.

These islands are included in the scene structure as references to the marked up scenes. At the moment, we only allow referring to formal scenes from informal text, which is enough to render the \Flyspeck text. Having an inclusion syntax fits the \Agora philosophy: the documentation workflow can use the formal code, but it should not change it. Instead, writing informal documentation about a development should be similar to writing a \LaTeX\ article, only in a different markup language. However, it is occasionally necessary to add code directly to an informal page, for example to write an illustrative example or a failed attempt; such a code block is not part of the formal development, but benefits from the markup techniques applied to the development.

In the document structure, such code blocks are just scenes, that are marked to be written in a particular language. From the rendered page, it is possible to open an editor for each scene, which requires special functionality to support writing formal proofs.

\section{Interaction with Formal \HOLLight Code}
\label{Interaction}
\subsection{Parsing and Proving}
For \HOLLight, adding \Proviola support implies adding a parser that can transform a proof script into a list of commands, and adding a layer to communicate with the prover's read-eval-print loop (REPL). This is sufficient, but so far does not create a very illustrative \Proviola display: most \HOLLight proofs are \emph{packaged} into a single REPL-invocation that introduces and discharges a theorem. Making this into a useful \Proviola display is left for future work, but we will sketch how a better display can be implemented using the scene structure of a \Proviola document.

To illustrate the workings of the parser and the prover, we use the following example code:

\begin{lstlisting}
(* Example code fragment. *)
g `x=x`;;
e REFL_TAC;;
let t = (* Use top_thm to verify the proof. *)
  top_thm();;
\end{lstlisting}

\paragraph{Parser} Because \HOLLight proofs are written as syntactically correct scripts that are interpreted by the \OCaml read-eval-print loop (REPL), the parser separates a proof script into the single commands that can be interpreted by this REPL. These commands are, in the \Flyspeck sources, terminated by `\texttt{;;}'\footnote{According to the \OCaml reference manual, \url{http://caml.inria.fr/pub/docs/manual-ocaml-4.00/manual003.html\#toc4}} and followed by a newline, so our parser splits a proof script into commands by looking for this terminator. 
Additionally, the proof can contain  comments, surrounded by `\texttt{(*}' and `\texttt{*)}': we let the parser only emit a command if the terminator does not occur as part of a comment. Finally, comment blocks that are not within other commands are treated as separate commands.  This last decision differs from traditional source-code parsers, which regard comments as white space, because \Agora reconstructs the proof script's appearance from the frames in the movie, in order to show the complete proof script if a reader desires it.

The parser does not group the frames into a scene structure: a \HOLLight proof is represented as a single scene containing all frames. For our example, the following frames are generated:
\begin{itemize}
\item \lstinline!(* Example code fragment. *)!
\item \lstinline!g `x=x`;;!
\item \lstinline!e REFL_TAC;;!
\item \lstinline!let t = (* Use top_thm to verify the proof. *)!\newline \lstinline!  top_thm ();;!
\end{itemize}

The first comment does not occur within a command, so it is parsed as a separate command, and the second comment occurs inside a command.

\paragraph{Prover}
\HOLLight is not implemented as a stand-alone program with its own REPL. Instead, it is implemented as a collection \OCaml scripts and some parsing functions. This means that the `prover' instance is actually a regular \OCaml REPL instance, which loads the appropriate bootstrap script. The problem of this approach is that these scripts take several minutes to load, a heavy penalty for wanting to edit a proof on the Web. 
To offset the load time, one can \emph{checkpoint} the \OCaml instance after it has bootstrapped \HOLLight. Checkpointing software allows the state of a process to be written to disk, and restore this state from the stored image later. We use \systemname{DMTCP}\footnote{\url{http://dmtcp.sourceforge.net}} as our checkpointing software: it does not require kernel modifications, and because of that is one of the few checkpointing solutions that works on recent Linux versions. 

Communication with the provers is encapsulated by a Python class: creating an instance of the class loads the checkpoint and connects to its standard input and output. The resulting object has a \texttt{send} method which writes a provided command to standard input and returns the REPL's response.
Beyond this low-level communication mechanism, the object also provides a \texttt{send\_frame} method. This method takes an entire frame and sends the command stored in it. This method does not only send the text, but also records the number of tactics that the prover has executed so far, by examining the length of the current goalstack. This gives an indication of how far a list of frames is processed, and allows the prover to use \HOLLight's undo function to prevent executing too many commands. 

After sending the frames generated from our example code, the frames have stack numbers as shown in Table~\ref{tab:frame-state}.

\begin{table}
\begin{center}
\begin{tabular}{l|l}
Command                                  & State \\\hline
\lstinline!(* Example code fragment. *)! & 0 \\
\lstinline!g `x=x`;;!                    & 1 \\
\lstinline!e REFL_TAC;;!                 & 2 \\
\lstinline!let t = ...!                  & 2
\end{tabular}
\end{center}
\caption{Frames with state numbers \label{tab:frame-state}}
\end{table}

When the frame with the \lstinline!REFL_TAC! invocation is changed, the \texttt{send\_frame} method will send the \HOLLight undo function, \lstinline!b ();;! as many times as is necessary to return to state 1. Afterwards, it will send the command of the changed frame.

The \HOLLight glue does not send all commands equally: the \Flyspeck formalization packs its proofs within an \OCaml module, which causes the REPL not to give output until the module is closed. Because we want to give state information per command, the gluing code ignores the \texttt{module} and \texttt{end} commands that signal the opening and closing of modules.

\paragraph{Packaged Proofs}
To allow \Proviola to record a packaged proof, it needs to break the proof down to its individual commands. To do this, we propose to use the \Tactician tool~\cite{AdamsAspinall-2012}: this is an extension to \HOLLight that records a packaged proof as it is executed, and allows the user to retrieve the actual tactics executed, which exposes the tree-like structure of such a proof: some of the tactics in the packaged proof might be applied multiple times, to different subgoals generated during the proof.

We can use the sequential tactic script generated by \Tactician directly, rendering it instead of the packaged proof, or do more sophisticated post-processing: we could match up the generated tactics to their occurrence in the packaged proof, and generate a special scene for each packaged proof. This scene would render as the original proof, but execute the \Tactician-generated sequence to provide responses. This gives readers a better feel of what is going on in such a packaged proof, but depends on a correct matching of the packaged proof to the sequential proof. We have not yet fully investigated the reach of these possibilities, however, so this remains as future work.

\subsection{Hyperlinking}
\label{Hyperlinking}
It seems that no proper hyperlinking facility exists so far for
\HOL-based systems. Such a facility should plug in to the parsing layer of the systems (as done, e.g., for \Coq
and \Mizar), and either export the information about symbols'
definitions relative to the original formal text, or directly produce
a hyperlinked version of the text: this hyperlinking pass should be fast, so it can be run when a page is loaded in the browser.

For \HOLLight (and \Flyspeck), we so far did not try to hook into the parsing layer of the system, and only provide a heuristic hyperlinking system. Still, such a
hyperlinker can be useful, because relatively few concepts are overloaded in the formalization, and most of the definitions and theorems are
introduced using a regular syntax: this means that the hyperlinker can generate an index for file definitions with only a small chance of ambiguity.
The hyperlinking proceeds in two broad steps, an indexing step and a rendering step. The indexing is done by a Perl script that generates a symbol index by:
\begin{enumerate}
\item collecting the globally defined symbols and theorem names from the formal texts by heuristically matching the most common patterns that introduce 
them,\footnote{To help this, we also use the theorem names stored by the HOL
 Light processing in the 
"theorems" file, using the mechanisms from the file
 update_database_**.ml.} 
and
\item optionally adding and removing some symbols based on a predefined list.
\end{enumerate}

The page renderer of \Agora then processes the texts again by heuristically tokenizing the text, looking up tokens and their linking in the generated index. Additionally, the page rendering also uses the index to generate metadata that can be used by the referencing mechanism~\cite{Tankink+-2012}.

The complete hyperlinking of the whole library now takes less than ten
seconds, and while obviously imperfect, it seems to be already quite
useful tool that allowed us to browse and study the library. The generated index of 15,780 \Flyspeck entities together with their URLs can be loaded into
arbitrary external application, and used for separate heuristic
hyperlinking of other texts. This function is used by the script that
translates the \LaTeX\ sources of the informal text describing \Flyspeck into wiki syntax
(Section~\ref{Creolifier}), to link the formally defined concepts to
their \HOLLight definitions.

\subsection{Editing and Proof Advising}
\begin{figure}
\includegraphics[width=\textwidth]{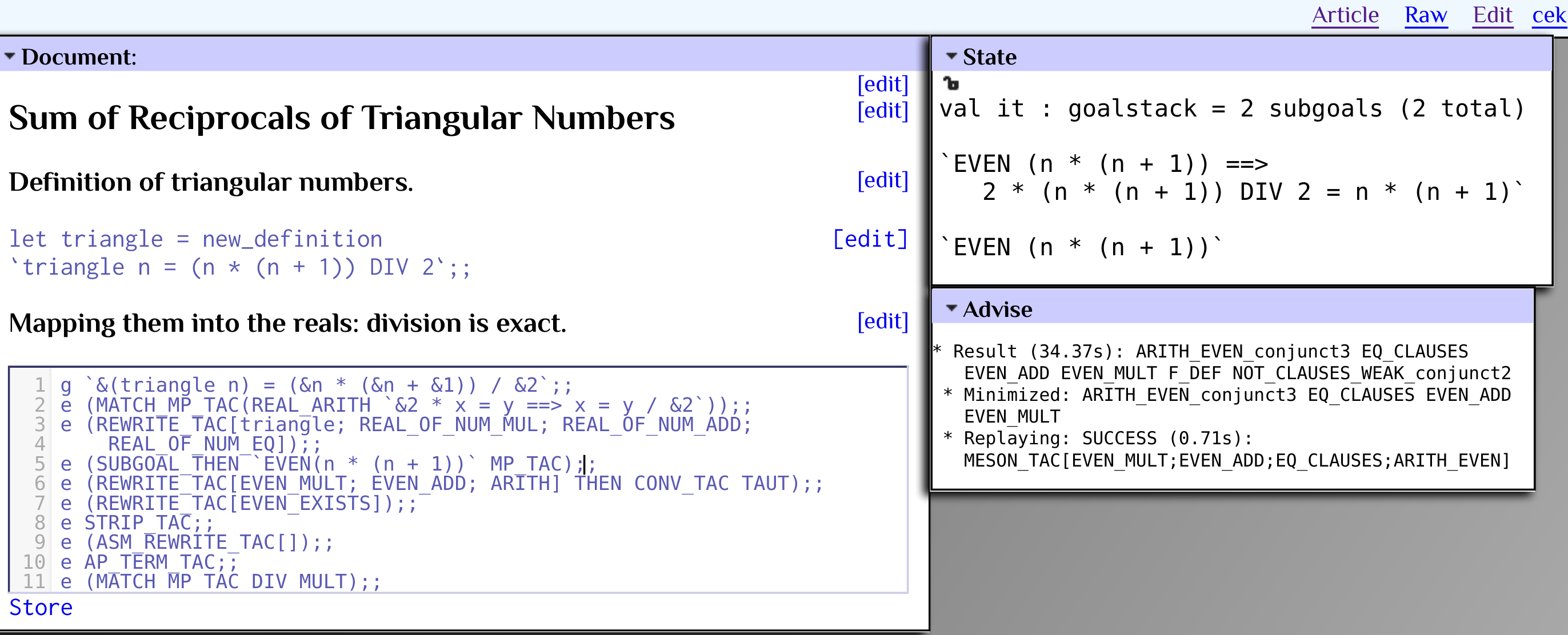}
\caption{The interactive editor built in the Wiki with the proof
state for the line with the cursor. The screenshot features a section
of Harrison's triangular numbers formalization. In line 5
the advisor automatically finds a proof that $n(n + 1)$ is even,
slightly different from the one used in the edited formalization.}
\label{fig:edit1}
\end{figure}

\paragraph{Editing} We can directly use the tools that turn text into frames for building the server backend of a (simple) web-based editor: the front end of this editor just gathers the entered text and sends it to the server, the server processes it into a list of frames and post-processes it: both by generating proof assistant (\HOLLight) responses and by sending markup information based on the correctness of a part of the text. Because this processing is incremental, information can be returned on demand: after the text has been parsed into frames, the server can give the editor information as it is produced, using the protocols described in~\cite{Tankink-2012}. As also described in that paper, it remains an open question on how to properly deal with the impact of the formal text written in the editor, as this might invalidate the entire repository. An example of the editor interaction is shown in Figure~\ref{fig:edit1}. It already shows also the proof advising facility.

\paragraph{Proof Advising}
In order to further facilitate the online Wiki authoring using
\HOLLight, we have added a post-processing step to the editor. For
each goal interactively computed by the proof assistant, the editor automatically
submits this goal to the AI/ATP proof advisor (\HH) service~\cite{hhemacs}. 
The advisor uses a number of differently
parametrized premise-selection methods (based on various
machine-learning algorithms) to find the most relevant theorems from
the \Flyspeck library for a given goal, and passes them (after translation to
first-order logic) to automated theorem provers (ATPs) such as
\Vampire~\cite{Vampire}, \E~\cite{Sch02-AICOMM}, and \Z~\cite{z3}.
If an ATP proof is found, it is minimized and reconstructed by a number of 
reconstruction strategies described in~\cite{hhrecon}.
In parallel to such AI/ATP methods, a number of decision
procedures are tried on the goal.
The currently used
decision procedures are able to solve boolean goals (tautologies), goals that involve
naturals (arithmetic), integers, rationals, reals and complex numbers including
Gr\"obner bases.
Whenever any of the strategies finds a tactic that solves the goal, all other
strategies are stopped and the result of the successful one is transmitted to the \Agora users through a window.  The users  can immediately use the
successful results in their proof.

The protocol to communicate with the advisor has been designed to be as
simple as possible, in order to enable using it not only as a part of
\Agora but also via an experimental \systemname{Emacs} interface~\cite{hhemacs} and
from the command line tool in the spirit of old style LCF. A request for
advice consists of a single line which is a text representation of a goal
to prove. To encode a goalstate as text the goal assumptions need to be
separated from the goal conclusion and from each other. We use the `
character as such separator, since the character never appears in normal
\HOLLight terms as it is used to denote start and end of terms by the
\systemname{Camlp5} preprocessor. 
When a request for advice is received the server parses the goal assumptions
and conclusion together, to allow matching the free variables present in
more than one of them and ensure proper typing.
The response is also textual and the connection is closed when no more advice for the goalstate is
available. Server-side caching is used to handle repeated queries,
typically produced by refactoring an existing proof script in the
Wiki.

\section{Inclusion of the Informal \Flyspeck Texts}
\label{Informal}
We have used a version of the informal \Flyspeck \LaTeX{} text that has 309 pages, but only a smaller part has so far been chosen for the experiments: Chapter 5 (Fan).
The file fan.tex has 1981 lines. There are 15 definitions (some of them define several concepts) and 36 lemmas. The definitions have the following annotated form (developed by Hales), which already cross-links to some of the formal counterparts (formally defined theorem names like \texttt{QSRHLXB} and \texttt{MUGGQUF} and symbols like \texttt{{azim\_fan}} and \texttt{is\_Moebius\_contour}):
\begin{small}
\begin{verbatim}
\begin{definition}[polyhedron]\guid{QSRHLXB}
A \newterm{polyhedron} is the
intersection of a finite number of closed half-spaces in
$\ring{R}^n$.  
\end{definition}
\end{verbatim}
\end{small}
The lemmas are written in a similar style:
\begin{small}
\begin{verbatim}
\begin{lemma}[Krein--Milman]\guid{MUGGQUF} 
Every compact convex set $P\subset\ring{R}^n$ is the convex hull 
of its set of extreme points.
\end{lemma}
\end{verbatim}
\end{small}
The text contains many mappings between informal and formal concepts, e.g.:
\begin{small}
\begin{verbatim}
\formaldef{$\op{azim}(x)$}{azim\_fan}
\formaldef{M\"obius contour}{is\_Moebius\_contour}
\formaldef{half space}{closed\_half\_space, open\_half\_space}
\end{verbatim}
\end{small}

\label{Creolifier}

There are several systems that can (to various extent) transform
\LaTeX{} texts to (X)HTML and similar formats. Examples include
LaTeXML\footnote{\url{http://dlmf.nist.gov/LaTeXML/}},
PlasTeX\footnote{\url{http://plastex.sourceforge.net/}},
xhtmlatex\footnote{\url{http://www.matapp.unimib.it/~ferrario/var/x.html}},
and TeX4ht.\footnote{\url{http://tug.org/tex4ht/}}  Often they are
customizable, and some of them can be equipped with custom non-HTML
(e.g., wiki) renderers. For the first experiments we have however
relied only on MathJaX for rendering mathematics, and custom
transformations from \LaTeX{} to wiki syntax that allow us to easily
experiment with specific functions for cross-linking and formalization
without involving the bigger systems. The price for this is that the
resulting wiki pages are more similar to presentations in ProofWiki
and Wikipedia than to full-fledged HTML book presentations. We might
switch to the larger extendable systems when it is clear what
extensions are needed for our use-case.

The transformations are now implemented in about 200 lines of a Perl
script (Creolify.pl) translating the \Flyspeck \LaTeX{} sources into the
enhanced Creole wiki syntax used by \Agora.  The script is easily
extendable, and it now consists mainly of about 30 regular-expression
replacements and related functions taking care of the
non-mathematical \LaTeX{} syntax and macros.  The mathematical text is
handled by the (slightly modified) macros taken from \Flyspeck
(kepmacros.tex) 
that are prepended to any \Agora \Flyspeck text and used automatically
by MathJax.  Producing and tuning the transformations took about one to two
days of work, and should not be a large time investment for (formal)
mathematicians interested in experimenting with \Agora.
The particular transformations that are now used for \Flyspeck include:

\begin{itemize}
\item Transformations that handle wiki-specific syntax that is
  (intentionally or accidentally) used in \LaTeX, such as
  comments, white space, fonts and section markup.
\item Transformations that create wiki subsections for various \LaTeX \ 
  blocks, sections, and environments. Each definition, lemma, remark,
  corollary, and proof environment gets its own wiki subsection,
  similarly, e.g., to ProofWiki and Wikipedia.
\item The transformation that add linking and cross-linking, based on
  the \LaTeX \  annotations. Each \LaTeX \ label produces a corresponding
  wiki anchor, and each \LaTeX \ reference produces a link to the
  anchor. Newly defined terms (introduced with the \texttt{newterm}
  macro) also produce anchors. Formal annotations (introduced with the
  \texttt{guid} and \texttt{formaldef} macros) are first looked up in
  the index of all formal concepts produced by hyperlinking of the
  formalization (Section~\ref{Hyperlinking}), and if they are found there, such annotations are
  linked to the corresponding formal definition.
\end{itemize}

\section{Conclusion and Future Work}
\label{Conclusion}
The platform is still in development, and a number of functions can be
improved and added. For example, whole-library editing, guarded by
global consistency checking of the formal code that has been already
verified (as done for \Mizar~\cite{UrbanARG10}), is future work.  On
the other hand, the platform already allows the dual presentation of
mathematical texts as both informal and formal, and the interaction
between these two aspects. 
In particular, the platform takes both \LaTeX{} and
formal input, cross-links both of them based on simple user-defined
macros and on the formal syntax, and allows one to easily browse the
formal counterparts of an informal text. 
It is already possible to add further
formal links to the informal concepts, and thus make the informal text
more and more explicit. A particular interesting use made possible by
the platform is thus an exhaustive collaborative formal annotation of the
\Flyspeck book. 
The platform also already includes interactive
editing and verification, 
which allows at any point
of the informal text to switch to formal mode, and to add the
 corresponding formal definitions, theorems, and proofs, which are immediatelly 
 hyperlinked and equipped with detailed proof status information for every step. 
The editing is complemented by a relatively strong proof
advice system for \HOLLight. This is especially useful in a Wiki
environment, where redundancies and deviations can be discovered
automatically. The requests for advice can become grounds for further
experiments on strengthening the advice system.

One future direction is to allow even the non-mathematical parts of
the wiki pages to be written directly with (extended) \LaTeX, as it is
done for example in PlanetMath. This could facilitate the presentation
of the projects developed in the wiki as standalone \LaTeX{} papers. On
the other hand, it is straightforward to provide a simple script that
translates the wiki syntax to \LaTeX, analogously to the existing
script that translates from \LaTeX{} to wiki.

\bibliographystyle{splncs}
\begin{small}
\bibliography{flyspeck}
\end{small}
\end{document}